\documentstyle[12pt]{article}

\catcode`@=11

\ifx\documentclass\undefined
 \def\@sect#1#2#3#4#5#6[#7]#8{\ifnum #2>\c@secnumdepth
     \let\@svsec\@empty\else
     \refstepcounter{#1}\edef\@svsec{\csname prefix#1\endcsname
        \csname the#1\endcsname\hskip 1em}\fi
     \@tempskipa #5\relax
      \ifdim \@tempskipa>\z@
        \begingroup #6\relax
          \@hangfrom{\hskip #3\relax\@svsec}{\interlinepenalty \@M #8\par}%
        \endgroup
       \csname #1mark\endcsname{#7}\addcontentsline
         {toc}{#1}{\ifnum #2>\c@secnumdepth \else
                      \protect\numberline{\csname the#1\endcsname}\fi
                    #7}\else
        \def\@svsechd{#6\hskip #3\relax  %% \relax added 2 May 90
                   \@svsec #8\csname #1mark\endcsname
                      {#7}\addcontentsline
                           {toc}{#1}{\ifnum #2>\c@secnumdepth \else
                             \protect\numberline{\csname the#1\endcsname}\fi
                       #7}}\fi
     \@xsect{#5}}
\else
    \def\@seccntformat#1{\csname prefix#1\endcsname
        \csname the#1\endcsname\quad}
\fi

\@addtoreset{equation}{section}   % Makes \section reset 'equation' counter.

\def\thebibliography#1{\section*{References\@mkboth
 {REFERENCES}{REFERENCES}}\list
 {\leftbibmark\arabic{enumi}\rightbibmark}{
 \settowidth\labelwidth{\leftbibmark #1\rightbibmark}\leftmargin\labelwidth
 \advance\leftmargin\labelsep
 \usecounter{enumi}}
 \def\newblock{\hskip .11em plus .33em minus -.07em}
 \sloppy\clubpenalty4000\widowpenalty4000
 \sfcode`\.=1000\relax}

\def\@citex[#1]#2{\if@filesw\immediate\write\@auxout{\string\citation{#2}}\fi
  \def\@citea{}\@cite{\@for\@citeb:=#2\do
    {\@citea\def\@citea{,\penalty\@m\ }\@ifundefined
       {b@\@citeb}{{\bf ?}\@warning
       {Citation `\@citeb' on page \thepage \space undefined}}%
\hbox{\csname b@\@citeb\endcsname\citemarkdelim}}}{#1}}

\def\@cite#1#2{\leftcitemark{#1\if@tempswa, #2\fi}\rightcitemark}
\def\leftcitemark{[}
\def\rightcitemark{]}
\def\citemarkdelim{}
\def\leftbibmark{[}
\def\rightbibmark{]}

\catcode`@=12

\def\A{{\cal A}}

\def\L{{\cal L}}

\def\R{{\cal R}}

\def\X{{\cal X}}

\def\linebreak{\hfill\break}

\def\Eq#1{Eq.(\ref{#1})}
\def\Eqs#1#2{Eqs.(\ref{#1})-(\ref{#2})}

\def\tend{\rightarrow}

\def\therefore{\mbox{\setbox0=\hbox{X}\hbox{$\ldotp$}\raise0.7\ht0\hbox{$\ldotp$}\hbox{$\ldotp$}} }
\def\because{\mbox{\setbox0=\hbox{X}\raise0.7\ht0\hbox{$\ldotp$}\hbox{$\ldotp$}\raise0.7\ht0\hbox{$\ldotp$}}\kern0pt }
\def\r#1{{\rm #1}}

\def\bm#1{\mbox{\boldmath $#1$}}

\def\Frac(#1/#2){\left(\frac{#1}{#2}\right)}

\def\bop{\mathchoice{{\scriptstyle\circ}}{{\scriptstyle\circ}}{{\scriptscriptstyle\circ}}{\circ}}

\def\Tdot#1{{{#1}^{\hbox{.}}}}

\def\In{\mathrel{\mbox{\setbox0=\hbox{$\cup$}\dimen0=\wd0\divide\dimen0 by 2
\box0\kern -\dimen0\vrule}}}

\def\Beq{\begin{equation}}
\def\Eeq{\end{equation}}
\def\Beqr{\begin{eqnarray}}
\def\Eeqr{\end{eqnarray}}
\def\Beqrn{\begin{eqnarray*}}
\def\Eeqrn{\end{eqnarray*}}
\def\Bitm{\begin{itemize}}
\def\Eitm{\end{itemize}}

\font\elevenmib=cmmib10 scaled\magstephalf   \skewchar\elevenmib='177

\begin{document}

\thispagestyle{empty}

\begin{titlepage}

%\hbox to \hsize{\YUKAWAmark \hfill YITP-97-62}
%\rightline{KUNS 1482}
%\rightline{December 1997}
%
%
%\vspace{2cm}

\begin{center}\large\bf
Evolution of Cosmological Perturbations \\
in the Long Wavelength Limit
\end{center}

\bigskip

\begin{center}
Hideo Kodama\footnote{email address: kodama@yukawa.kyoto-u.ac.jp} 
\end{center}

\begin{center}\it
Yukawa Institute for Theoretical Physics, Kyoto University, \\
Kyoto 606-01, Japan\\
\end{center}

\begin{center}
and
\end{center}

\begin{center}
Takashi Hamazaki
\footnote{email address: hamazaki@murasaki.scphys.kyoto-u.ac.jp}
\end{center}

\begin{center}\it
Department of Physics, Faculty of Science, Kyoto University,\\
Kyoto 606-01, Japan
\end{center}

\bigskip
\bigskip
\begin{center}\bf Abstract\end{center}
The relation between the long wavelength limit of solutions to the
cosmological perturbation equations and the perturbations of solutions
to the exactly homogeneous background equations is investigated for
scalar perturbations on spatially flat cosmological models. It is
shown that a homogeneous perturbation coincides with the long
wavelength limit of some inhomogeneous perturbation only when the
former satisfies an additional condition corresponding to the momentum
constraint if the matter consists only of scalar fields. In contrast,
no such constraint appears if the fundamental variables describing the
matter contain a vector field as in the case of a fluid. Further, as a
byproduct of this general analysis, it is shown that there exist two
universal exact solutions to the perturbation equations in the long
wavelength limit, which are expressed only in terms of the background
quantities. They represent adiabatic growing and decaying modes, and
correspond to the well-known exact solutions for perfect fluid systems
and scalar field systems.

\end{titlepage}

\section{Introduction}

In the current standard scenario based on the gravitational
instability theory, the present large scale structures of the universe
are formed through the following four stages: production of seed
fluctuations in the early universe, their linear evolution on
superhorizon scales, the subsequent linear modulation after they enter
the horizon, and the final non-linear evolutions. In the inflationary
universe models the seed fluctuations in the first stage are produced
from quantum fluctuations on the Hubble horizon scales during
inflation, and we now have universal formulae to determine their nature 
such as the amplitudes and the spectrum for a wide variety of
inflation models. We can now also easily handle the evolution during
the third stage because the matter content of the universe during this 
stage is rather simple and restricted. Of course the actual behavior
of perturbations on subhorizon scales are quite complicated even in
the linear stage, and we need numerical computations to determine
their details. 

On the other hand universal formulae to determine the evolution of
perturbations during the second stage have not been established yet,
although it is generally believed that the so-called Bardeen parameter
is conserved with good accuracy during this stage and this conservation
law essentially determines the amplitudes and the spectrum of
perturbations when they reenter the horizon, from those at the first
stage\cite{Bardeen.J&Steinhardt&Turner1983}.  Of course the
conservation of the Bardeen parameter during the Friedmann stage is
well established including the case in which the equation of state of
cosmic matter changes slowly
\cite{Kodama.H&Sasaki1984,Hamazaki.T1997}. The simplicity comes
from that fact that the evolution of perturbations on superhorizon
scales are determined with good accuracy by that on the long
wavelength limit
\cite{Mukhanov.V1988,Polarski.D&Starobinsky1992,Kodama.H&Hamazaki1996}.
In this limit
the evolution of adiabatic modes is determined by a set of simple
1st-order ordinary differential equations for two gauge-invariant
variables, from which follows the exact conservation of the Bardeen
parameter \cite{Kodama.H&Sasaki1984,Hamazaki.T1997}.  However,
for some important situations, such as the reheating stage of
inflation models, this simple analysis does not work because the
adiabatic mode may produce entropy modes which feed back to the
behavior of the adiabatic modes\cite{Hamazaki.T&Kodama1996}.  In these
situations the knowledge on the evolution of perturbations of all the
components of matter in the long wavelength limit is required to
determine the behavior of the Bardeen parameter.

In this connection Nambu and Taruya recently wrote an interesting
paper \cite{Taruya.A&Nambu1997} in which they stated that solutions to
the gauge-invariant perturbation equations for a multi-component
scalar field system on expanding universe in the long wavelength limit
are obtained as derivatives of exactly homogeneous solutions to the
Einstein equations with respect to the solution parameters.  If this
result is correct and all the solutions to the perturbation equations
can be obtained in this way, it provides a very powerful tool to
analyze the behavior of perturbations during inflation and reheating
including the problem of the conservation of the Bardeen
parameter. However, no proof nor explanation on their statement was
given in their paper. Further, direct calculation shows that the
perturbation equations in the long wavelength limit do not coincide
with those for exactly homogeneous perturbations.

In order to see whether Nambu and Taruya's statement is true or not as
well as to obtain a deeper understanding on the behavior of
perturbations on superhorizon scale perturbations, in the present
paper, we investigate the relation between the solutions to the
perturbation equations in long wavelength limit and the exactly
homogeneous solutions to the Einstein equations for multi-component
systems on spatially flat Robertson-Walker universe. Our arguments are
quite general except for the assumptions that the fundamental
variables describing matter are scalar fields and/or vector fields,
and that the amplitude of anisotropic stress perturbations vanishes
rapidly enough in the long wavelength limit. We will clarify general
conditions under which the perturbation solutions in the long
wavelength limit coincide with some exactly homogeneous perturbations
to the homogeneous solutions to the Einstein equations. In particular,
we will show that for the multi-component scalar field system not all
the exactly homogeneous perturbations directly correspond to perturbation
solutions in the long wavelength limit. Further, as a byproduct of our
analysis, we will show that the perturbation equations in the long
wavelength limit have two universal adiabatic solutions which can be
expressed explicitly as time integrals of known background quantities.

The paper is organized as follows. First in the next section we give
basic definitions of perturbation variables used in the present paper
and the perturbation equations for them. By inspecting the dependence
of these quantities and equations on the wave number $k$ of
perturbations, in \S3 we derive the conditions on  the exactly homogeneous
perturbations to coincide with the $k\tend0$ limit
of some solutions to the perturbation equations with $k\not=0$.
In \S4, as a special application of the argument in \S3, we show the 
existence of two universal adiabatic modes in the $k\tend0$
limit. Then in \S5 and \S6 we specify our arguments to a multi-component
scalar field system and a multi-component perfect fluid system,
respectively, and clarify the relation between the perturbations 
in the $k\tend0$ limit and those with $k=0$. Section 7 is devoted to
summary and discussions.

\section{Perturbation Equations}

In this section we recapitulate the definitions of basic perturbation
variables and their equations in the framework of the gauge-invariant
perturbation theory.  We adopt the notations used in the review article 
by Kodama and Sasaki\cite{Kodama.H&Sasaki1984}.

We only consider perturbations on a spatially flat($K=0$) Robertson-Walker 
universe throughout the paper. Hence the background metric is given by
\Beq
ds^2 = -dt^2 + a(t)^2 d\bm{x}^2,
\Eeq
and its perturbation by
\Beq
d\tilde s^2 = -(1+2A Y)dt^2 -2aB Y_j dt dx^j 
+ a^2\left[(1+2H_LY)\delta_{jk}+ 2H_T Y_{jk}\right]dx^j dx^k,
\label{MetricPerturbation}\Eeq
where $Y$, $Y_j$ and $Y_{jk}$ are harmonic scalar, vector and tensor
for a scalar perturbation with wave vector $\bm{k}$ on flat 3-space:
\Beq
Y :=e^{i k\cdot x},\quad
Y_j :=-i{k_j\over k}Y,\quad
Y_{jk} :=\left({1\over3}\delta_{jk}-{k_jk_k\over k^2}\right)Y.
\Eeq

Under the infinitesimal gauge transformation
\Beq
\delta t=T Y,\quad \delta x^j=L Y^j,
\Eeq
the metric perturbation variables in \Eq{MetricPerturbation} transform as
\Beqr
&& \bar A=A -\dot T,\\
&& \bar B=B +a\dot L + {k\over a}T,\\
&& \bar H_L=H_L -{k\over3}L - HT,\\
&& \bar H_T=H_T +kL
\Eeqr
where the overdot represents the derivative with respect to the proper time
$t$ and $H$ is the cosmic expansion rate $\dot a/a$. From this it follows
that the spatial curvature perturbation $\R$ 
and the shear $\sigma_g$ defined by
\Beqr
&& \R := H_L + {1\over3}H_T,\\
&& \sigma_g := {a\over k}\dot H_T -B
\Eeqr
transform as
\Beq
\bar\R =\R -HT,\quad
\bar\sigma_g =\sigma_g -{k\over a}T.
\Eeq
Hence we obtain the following two independent gauge-invariant combinations:
\Beqr
&& \A := A-\Tdot{(\R / H)},\\
&& \Phi := \R - {aH\over k}\sigma_g.
\Eeqr
All the other gauge-invariant combinations of metric perturbation variables are
written as linear combinations of $\A$ and $\Phi$.

Here note that for the exactly homogeneous perturbations corresponding to $k=0$,
the vector-like and the tensor-like perturbation variables should vanish. 
Hence for the metric perturbation, $B=H_T=0$. Correspondingly, the gauge
freedom for them is just the time reparametrization $T$, and $L$ should
vanish in the gauge transformation. Therefore there exists only one 
gauge-invariant combination for the exactly homogeneous metric perturbations,
which coincides with $\A$. The variable $\Phi$ has no counterpart for them.
This point is very important in the argument on the correspondence between 
the perturbations in the $k\tend 0$ limit and the exactly homogeneous ones.

Next we consider the matter perturbations. In this paper we  assume
that the fundamental variables describing matter are scalar
and vector quantities. This assumption is satisfied in most of the 
realistic applications. Let us denote the corresponding perturbation 
variables by $\chi_I$ and $v_P$ where $I$ and $P$ are indices labeling 
components. Then they transform under the gauge transformation as
\Beq
\bar\chi_I=\chi_I -\dot S_I T,\quad
\bar v_P = v_P + a\dot L,
\Eeq
where $S_I$ is the background quantity corresponding to the perturbation
$\chi_I$. Hence the corresponding gauge-invariant variables are given by
\Beqr
&& X_I:= \chi_I - {\dot S_I\over H}\R,\\
&& V_P:= v_P - {a\over k}\dot H_T.
\Eeqr

In the exactly homogeneous system the matter is described only by the
scalar quantities $S_I$. Hence its perturbation gives only 
the scalar-type gauge-invariant variables $X_I$ as in the case of
the metric perturbation.

The Einstein equations relate these matter variables with the metric 
variables through the energy-momentum tensor. For scalar perturbations
its generic form is written as
\Beqr
&& \tilde T^0_0=-(\rho+\delta\rho Y),\\
&& \tilde T^0_j = a(\rho+p)(v-B)Y_j,\\
&& \tilde T^j_k = (p\delta^j_k + \delta p\delta^j_k + \Pi Y^j_k),
\Eeqr
where $\rho$ and $p$ are the background values of the energy density and
the pressure, respectively, and follow the equations,
\Beqr
&& H^2={\kappa^2\over3}\rho,\label{HubbleEq}\\
&&\dot \rho=-3(\rho+p)H \label{EnergyEq}.
\Eeqr
In the present paper we only consider the case in which the anisotropic 
stress perturbation $\Pi$, which is gauge-invariant by itself,  
vanishes faster than $k^2$ in the $k\tend 0$ limit.

Applying the general argument on the matter perturbation above, we can
construct from the density perturbation $\delta\rho$, the velocity perturbation
$v$ and the isotropic stress perturbation $\delta p$ the following
three gauge-invariant combinations:
\Beqr
&& \rho \Delta_g:= \delta\rho + 3(\rho+p)\R,\\
&& V:=v-{a\over k}\dot H_T,\\
&& p\Gamma:=\delta p - c_s^2\delta \rho,
\Eeqr
where $c_s^2:=\dot p/\dot \rho$ is the square of the sound velocity.

Now in terms of these gauge-invariant variables the perturbation of
the Einstein equations are written as
\Beqr
&\tilde G^0_0=\kappa^2 \tilde T^0_0:
& \A + {1\over2}\Delta_g ={k^2\over 3a^2H^2}\Phi,
\label{HamiltonianConstraint}\\
&\tilde G^0_j=\kappa^2 \tilde T^0_j:
& k\left[\A+ {3\over2}(1+w)Z\right]=0,
\label{MomentumConstraint}\\
&\tilde G^j_j=\kappa^2 \tilde T^j_j:
& H\dot\A + (2\dot H+3 H^2)\A={\kappa^2\over2}(p\Gamma+c_s^2\rho\Delta_g)
-{\kappa^2\over3}\Pi,
\label{PertEnergyEq}\\
&\tilde G^j_k-\tilde G^l_l\delta^j_k
=\kappa^2 (\tilde T^j_k-\tilde T^l_l\delta^j_k):
& k^2\left[\A + {1\over a}\Tdot{\left({a\over H}\Phi\right)}\right]
=-\kappa^2 a^2\Pi,\label{PotentialEq}
\Eeqr
where $w=p/\rho$ and $Z$ is the Bardeen parameter defined by
\Beq
Z:=\Phi-{aH\over k}V=\R - {aH\over k}(v-B).
\Eeq
If we eliminate $\Delta_g$ from \Eq{PertEnergyEq} using 
\Eq{HamiltonianConstraint}, we obtain
\Beq
\Tdot{\left(\A\over 1+w\right)}={c_s^2\over 1+w}{1\over H}{k^2\over a^2}\Phi
+{H\over \rho +p}\left({3\over2}p\Gamma-\Pi\right).
\label{PertEnergyEq1}\Eeq

Note that we have not multiplied any power of $k$ in deriving these equations
from the Einstein equations. Hence putting $k=0$ in these equations 
yields the equations for the exactly homogeneous perturbations 
if one takes account of the fact that the terms multiplied by inverse 
powers of $k$ vanishes identically  for these perturbations as mentioned 
above.

In general these equations must be supplemented by the expressions for
$\Delta_g$, $Z$, $\Gamma$ and $\Pi$ in terms of the fundamental gauge-invariant
matter variables $X_I$ and $V_P$ and their evolution equations. We do not 
write them explicitly here because the details of these equations are not
necessary until we specialize the general arguments to specific models.

\section{$k=0$ vs $k\tend0$ limit}

In this section we clarify under what conditions the $k\tend0$ limit
of solutions to \Eqs{HamiltonianConstraint}{PotentialEq} 
coincide with some solutions to the corresponding equations for $k=0$.
For that purpose first note that the gauge-invariant variables
introduced in the previous section are classified into two groups.

The first one consists of the gauge-invariant variables whose expressions
in terms of the gauge-variant perturbation variables do not contain
inverse powers of $k$. $\A$, $\Delta_g$, $\Gamma$ and $X_I$ belong to
this group. These variables always have well-defined finite 
$k\tend0$ limits which
coincide with the corresponding gauge-invariant variables for the 
exactly homogeneous perturbations. Of course in taking $k\tend0$ limit
the conditions
\Beq
B, H_T, v_P, v, \Pi/k^2 \tend 0
\label{AsymptCond:general}\Eeq
should be taken into account.

On the other hand the second group consists of the gauge-invariant variable
which contain inverse powers of $k$ when expressed in terms of gauge-variant
variables. $\Phi$, $Z$(or $V$) and $V_P$ belong to this group. These
variables have no counterpart for the exactly homogeneous perturbations,
and their $k\tend0$ limits depend on how fast the quantities in 
\Eq{AsymptCond:general} vanish in the limit.

With these points in mind, let us first consider the $k\tend0$ limit of
the perturbation of the equations for the fundamental matter variables.
When expressed in terms of the original gauge-variant perturbation
variables $\chi_I$, $v_P$, $A$, $B$, $H_L$ and $H_T$, they do not
contain inverse powers of $k$. We can write them in terms of gauge-invariant
variables in the following way. First eliminate $\chi_I$ and $A$ by replacing
them by $X_I$, $\A$ and $\R$.  Second eliminate $v_P$ using $V_{PQ}=v_P-v_Q$  
and $Z$. These procedures do not produce terms with inverse
powers of $k$, and the coefficient of $Z$ contains a positive power of $k$.
Now since all the gauge-variant matter variables and $A$ are eliminated,
the remaining terms should be proportional to $\Phi$( and its time derivatives).
This implies that in the final expressions $Z$ and $\Phi$ appears only 
in the forms $k^mZ$($m\ge1$) and $k^n\Phi$($n\ge2$). Further the other
terms, which are proportional to $\A$, $X_I$ or $V_{PQ}$, 
contain no inverse power of $k$. Therefore, taking into account the
condition (\ref{AsymptCond:general}), the $k\tend0$ limit of a solution
to the perturbation equations satisfies the corresponding equations
for $k=0$ if and only if
\Beqr
&& k^2\Phi\tend 0,\label{AsymptCond1}\\
&& kZ \tend 0,\label{AsymptCond2}\\
&& V_{PQ} \tend 0.\label{AsymptCond3}
\Eeqr

Next let us examine the Einstein equations. Among the four equations
it is easy to see that \Eq{HamiltonianConstraint} and 
\Eq{PertEnergyEq} reduce to the perturbation of \Eq{HubbleEq}
and \Eq{EnergyEq} in the $k\tend0$ limit under the condition
$k^2\Phi\tend0$. On the other hand 
\Eq{MomentumConstraint} and \Eq{PotentialEq} become trivial 
for the exactly homogeneous perturbations. However, for $k\not=0$,
these equations yield non-trivial perturbation equations
\Beqr
&& \A + {3\over2}(1+w)Z=0, \label{AdditionalConstraint}\\
&& \A + {1\over a}\Tdot{\left({a\over H}\Phi\right)}=0,
\label{PotentialEq1}
\Eeqr
where we have put $\Pi=0$ because it does not affect the arguments on
the $k\tend0$ limit under the assumption $\Pi/k^2\tend0$ adopted in the
present paper.

The second equation of these can be solved in terms of $\Phi$ as
\Beq
\Phi={H\over a}\left(C - \int_{t_0} a(t)\A(t) dt\right),
\label{PhiByA}\Eeq
where $C$ is an integration constant and $t_0$ is an initial time.
This equation can be regarded as an equation to determine the 
$k\tend0$ limit of $\Phi$ from a solution for $\A$ to the 
perturbation equations with $k=0$. In this viewpoint the consistency
condition (\ref{AsymptCond1}) is simply replaced by the condition
on the $k$-dependence of $C$,
\Beq
k^2C(k) \tend 0.
\Eeq
On the other hand the condition (\ref{AsymptCond2}) is always satisfied 
under \Eq{AdditionalConstraint}. Further the condition (\ref{AsymptCond3}) 
just select a subset of solutions to the perturbation equations with $k\not=0$,
and does not give any restriction on the exactly homogeneous perturbations.
Therefore starting from any solution to the
perturbation equations with $k=0$, one can always construct the 
gauge-invariant quantities representing the $k\tend0$ limit of a 
solution to the perturbation equations with $k\not=0$ by supplementing
$\A$ and $X_I$ for the exactly homogeneous perturbation with $\Phi$ 
determined by \Eq{PhiByA}, provided that \Eq{AdditionalConstraint}
do not yield any additional constraint on the 
seed exactly homogeneous perturbation.

In the case in which the fundamental variables describing matter
contains a dynamical vector field, the $k\tend0$ limit of $Z$ depends
on the value $\lim_{k\tend0}(v-B)/k$ which cannot be determined from the
information of the exactly homogeneous perturbations. Hence 
\Eq{AdditionalConstraint} can be simply regarded as the equation to 
determine the $k\tend0$ limit of $Z$. 

In contrast, in the case in which the matter is described only by 
scalar fields, $v-B$ in $\tilde T^0_j$ should be written as a 
combination of the spatial derivatives of $\chi_I$. Hence it must be
proportional to a positive power of $k$, and $Z$ has a well-defined
$k\tend0$ limit which is written only in terms of $X_I$. Thus 
the condition \Eq{AdditionalConstraint} yields a restriction on 
the seed exactly homogeneous perturbation in order for it to be 
a $k\tend0$ limit of some solution to the perturbation equations
with $k\not=0$.

\section{Universal Adiabatic Solutions}

Let $a(t)$ and $S_I(t)$ be an exactly homogeneous solution describing
the background. Then, since the scale factor $a(t)$ comes into the
evolution equations only through $H=\dot a/a$, $(1+\lambda)a(t)$
and $S_I(t)$ also gives an exactly homogeneous solution to the 
Einstein equation where $\lambda$ is a constant. If we regard this
solution as an exactly homogeneous perturbation, all the corresponding
gauge-variant perturbation variables for matter vanish, and the metric
perturbation is given by
\Beq
A=B=0,\quad \R=H_L=\lambda.
\label{ScaleTrans}
\Eeq
 From this we obtain
\Beq
\A=-\Tdot{\left(\R\over H\right)}=-{3\over2}(1+w)\lambda.
\label{AdiabaticMode:A}\Eeq
Equation (\ref{PhiByA}) determines $\Phi$ from this as
\Beq
\Phi=C{H\over a} + {3\over2}\lambda {H\over a}\int_{t_0}(1+w)a(t)dt.
\label{AdiabaticMode:Phi}
\Eeq

If there exists a dynamical vector field describing matter, this
equation and \Eq{AdditionalConstraint} determines $Z$ as
\Beq
Z=-{2\over 3(1+w)}\A=\lambda.
\label{AdiabaticMode:Z}\Eeq
On the other hand, if the matter is described only by scalar fields,
$v-B$ is written as $v-B=k\chi$ where $\chi$ is a combination of 
the scalar field perturbations $\chi_I$ which does not contain
a negative power of $k$. Since the matter perturbation vanishes
in the present case, this term should vanish, which implies that
$Z=\lambda-aH\chi=\lambda$. Hence the condition (\ref{AdditionalConstraint})
is satisfied.

Thus we find that the perturbation equations have always two solution
for which the $k\tend0$ limits of $\A$, $\Phi$ and $Z$ are given by
\Eqs{AdiabaticMode:A}{AdiabaticMode:Z}. Since the matter is not
perturbed, $\Gamma=0$ for them. Hence they represent adiabatic modes
in the $k\tend0$ limit. Clearly the solution proportional to $C$ is a
decaying mode, while that proportional to $\lambda$ is a growing mode
because the Bardeen parameter $Z$ is a non-vanishing constant.
Note that these solutions are universal in the sense that they are
valid for any matter contents and interactions.

This universal solutions cover most of the exact solutions in the
$k\tend0$ limit found so far, and the general argument developed here
explains why such exact solutions were found for large variety of 
matter content. For example, for the case in which matter consists of
a single-component perfect fluid, there exist two adiabatic modes.
The above solutions just give their $k\tend0$ limit, and in terms of
the standard variables $\Delta=\delta\rho/\rho+3(1+w)aH(v-B)/k$ and 
$V$ they are expressed as
\Beqr
&& \Delta={2k^2\over3 a^2H^2}\Phi
={2k^2\over 3Ha^3}\left[C+{3\over 2}\lambda\int_{t_0}(1+w)a dt\right],\\
&& V={k\over aH}(\Phi-Z)={k\over a^2}\left[C-\lambda
\left(a(t_0)/H(t_0) + \int_{t_0}a dt\right)\right],
\Eeqr
which recover the well-known exact solutions\cite{Kodama.H&Sasaki1984}. 

Another example is the case
in which matter consists of a multi-component scalar field $\phi=(\phi_I)$.
In terms of the gauge-invariant variable for the scalar field perturbation
defined by
\Beq
X_I= \delta\phi_I - {\dot\phi_I\over H}\R,
\label{X:scalar:def}
\Eeq
the above solution gives
\Beq
X_I=-{\dot\phi_I\over H}\lambda,
\Eeq
and 
\Beq
\Phi=C{H\over a} + \lambda{\kappa^2\over2}{H\over a}
\int {a\dot\phi^2\over H^2}dt.
\Eeq
These give the extension of the well-known exact solution in the $k\tend0$ 
limit for the single component scalar field 
case\cite{Polarski.D&Starobinsky1992,Kodama.H&Hamazaki1996,Mukhanov.V&Feldman&Brandenberger1992}.

\section{Scalar Field Systems}

In this section we examine how the condition (\ref{AdditionalConstraint})
is expressed explicitly for a multi-component scalar field system.

The Lagrangian density for a multi-component scalar $\phi=(\phi_I)$
is generally expressed as
\Beq
\L=-\sqrt{-g}\left[{1\over2}g^{\mu\nu}\partial_\mu\phi\cdot
\partial_\nu\phi + U(\phi)\right],
\label{Lagrangian:scalar}\Eeq
where $U(\phi)$ is a potential. For a homogeneous background the 
energy density $\rho$ and the pressure $p$ are expressed as
\Beq
\rho={1\over2}\dot\phi^2 + U(\phi),\quad
p={1\over2}\dot\phi^2-U(\phi),
\Eeq
where $\dot\phi^2=\dot\phi\cdot\dot\phi$. The background equation of motion
and the Einstein equations are given by
\Beq
\ddot \phi + 3H\dot\phi + D U=0,
\label{FieldEq:Background}\Eeq
where $DU=(\partial U/\partial \phi_I)$, and \Eq{HubbleEq}.

Since the energy-momentum tensor for the Lagrangian density 
(\ref{Lagrangian:scalar}) is given by
\Beq
T^\mu_\nu=\nabla^\mu\phi\cdot\partial_\nu\phi
-{1\over2}\delta^\mu_\nu\left(\nabla^\lambda\phi\cdot\nabla_\lambda + 2U
\right),
\Eeq
the perturbation variables for the energy-momentum tensor are expressed as
\Beqr
&& \delta\rho=-A\dot\phi^2 + \dot\phi\dot\delta\phi + DU\cdot\delta\phi,\\
&& (\rho+p)(v-B)={k\over a}\dot\phi\cdot\delta\phi,\\
&& \delta p = -A\dot\phi^2 + \dot\phi\dot\delta\phi - DU\cdot\delta\phi.
\Eeqr
Hence $\Delta_g$ and $Z$ are expressed in terms of the gauge-invariant
variable $X$ for the scalar field perturbation defined by \Eq{X:scalar:def}
as
\Beqr
&& \rho\Delta_g = -\A\dot\phi^2 + \dot\phi\cdot\dot X + DU\cdot X,\\
&& Z=-H{\dot\phi\cdot X \over \dot\phi^2}.
\Eeqr

Inserting this expression for $\Delta_g$ into \Eq{HamiltonianConstraint},
we obtain
\Beq
2U\A + \dot\phi\cdot\dot X + DU\cdot X=
2{1\over \kappa^2}{k^2\over a^2}\Phi.
\Eeq
 From this, the expression for $Z$ and the background field equation,
it follows that
\Beq
\A+{3\over2}(1+w)Z=-{H\over 2U}W\left({\dot\phi\over H},X\right)
+{1 \over \kappa^2 U}{k^2\over a^2}\Phi,
\Eeq
where
\Beq
W(X_1,X_2):=X_1\cdot \dot X_2 - \dot X_1\cdot X_2.
\Eeq
Hence the condition (\ref{AdditionalConstraint}) is expressed as
\Beq
\lim_{k\tend0}W\left({\dot\phi\over H},X\right)=0.
\label{AdditionalConstraint:scalar}\Eeq

Now let us see what kind of restriction this condition gives for
exactly homogeneous perturbations. Let us denote the gauge-invariant
variable for an exactly homogeneous perturbation $\delta \phi$ of
the scalar field as $\X=\delta\phi - \R\dot\phi/H$, where
\Beq
\R=H_L=\delta a/a.
\Eeq
Then from the perturbation of \Eq{FieldEq:Background} and \Eq{HubbleEq},
we obtain
\Beqr
&& \ddot\X + 3H\dot\X + D^2U(\X)-\dot\phi \dot \A + 2DU \A=0,\\
&& 2U\A + \dot\phi\cdot\dot\X + DU\cdot\X=0,
\Eeqr
where $D^2U(\X)=(\X_J\partial^2 U/\partial\phi_I\partial\phi_J)$.
Here note that $\X=\dot\phi/H$ and $\A=3\dot\phi^2/(3\rho)=3(1+w)/2$
is a solution to this equation, which corresponds to the adiabatic 
growing mode obtained in the previous section.

Eliminating $\A$ from these equations we obtain the following
second-order differential equation for $\X$:
\Beq
L(\X)=-{H^2\over U}\Tdot{\left({\dot\phi\over H}\right)}
W\left({\dot\phi\over H},\X\right),
\label{k0eq:scalar}\Eeq
where 
\Beq
L(\X):=\ddot\X + 3H\dot\X + \left[D^2U 
-{\kappa^2 \over a^3} \Tdot{\left( {a^3 \over H} \dot\phi \bop \dot\phi  
\right)}
%\dot\phi\bop\Tdot{\left(\dot\phi\over H\right)}
%+\kappa^2DU\bop{\dot\phi\over H}
\right]\X.
\label{L:def}\Eeq
 From this it follows that $W=W(\dot\phi/H,\X)$ follows the 
differential equation
\Beq
\dot W -\left(3wH + {1\over U}DU\cdot\dot\phi\right)W=0.
\label{Eq:W}\Eeq

This equation shows that if $W$ vanishes at an initial time $t=t_0$, it
vanishes at any time. Hence \Eq{AdditionalConstraint:scalar}
reduces to the condition on the initial value of solutions to
\Eq{k0eq:scalar}. Therefore, taking into account the fact that
$\X=\dot\phi/H$ is one of such solutions, we find that for a $N$-component
scalar field system $(2N-1)$ independent solutions to \Eq{k0eq:scalar}
correspond to the $k\tend0$ limit of solutions to the perturbation
equation with $k\not=0$. 

Here note that we have already obtained two universal solutions in the
$k\tend0$ limit in the previous section. Among them, the growing
mode, coincides with one of the solutions obtained from $\X$. On the
other hand the decaying mode is not contained in the latter because
$\X=0$ ($\delta a/a=0$) for it. Hence we have obtained 
$2N$ independent solutions in the $k\tend0$ limit. Since the number of
the dynamical degrees of freedom of the $N$-component scalar field system
is $2N$, these exhaust all the solutions to the perturbation equations
in the $k\tend0$ limit. However, if one wants to determine the 
$k$-dependence of the solutions around $k=0$ beyond the $k\tend0$ limit
by solving the perturbation equations iteratively with respect to $k$,
one needs all the solutions to the perturbation equation for $X$
in the $k\tend0$ limit to find the Green function. 
Fortunately we can obtain the remaining one independent solution for
$X$ explicitly from a solution for $\X$ with $W\not=0$ in the following
way\footnote{This point is first suggested out by Sasaki and Tanaka from
the consideration on the role of the $k\tend0$ limit of $H_T$ in the
discrepancy between the equations for $X$ and $\X$(private communication)}.

First note that the perturbation equation for $X$ with $k\not=0$ 
is expressed in terms of the operator $L$ defined in \Eq{L:def} as
\Beq
L(X) + {k^2\over a^2}X=0.
\Eeq
 From this we see that in the $k\tend0$ limit $X$ satisfies the
equation $L(X)=0$, which confirms the above argument. 

On the other hand from \Eq{Eq:W} it follows that
\Beq
{a^3H^2\over U}W\left({\dot\phi\over H},\X\right)={\r const}.
\Eeq
 From this find that 
\Beq
L(f\dot\phi/H)={\dot\phi\over a^3 H}\Tdot{(a^3\dot f)}
+2\dot f\Tdot{\left(\dot\phi\over H\right)}
\Eeq
coincides with the right-hand side of \Eq{k0eq:scalar} if
$f$ satisfies
\Beq
a^3\dot f=-{a^3H^2\over 2U}W\left({\dot\phi\over H},\X\right).
\Eeq
This implies that
\Beq
X=\X + {\dot\phi\over H}\int dt{H^2\over 2U}W\left({\dot\phi\over H},\X\right)
\label{X:GeneralSolution}\Eeq
satisfies $L(X)=0$ for any solution $\X$ to \Eq{k0eq:scalar}.
Since the second term on the right-hand side of this equation vanishes
if $W(\dot\phi/H,\X)=0$, and $W(\dot\phi/H,X)$ does not vanishes
if $W(\dot\phi/H,\X)\not=0$ from
\Beq
W\left({\dot\phi\over H},X\right)={\rho\over U}W\left({\dot\phi\over H},\X\right),
\Eeq
$X$ given by \Eq{X:GeneralSolution} exhausts all the solutions to 
the perturbation equation for $X$ in the $k\tend0$ limit.

\section{Perfect Fluid Systems}

In this section we apply the argument in \S3 to a multi-component
perfect fluid system as an example of non-trivial systems in which 
matter variables contain a dynamical vector field. 

The equations of motion of a perturbed multi-component system are
given by
\Beq
\tilde\nabla_\nu \tilde T^\nu_{I \mu}=\tilde Q_{I \mu}
\equiv \tilde Q_I \tilde u_\mu +\tilde f_{I\mu},
\label{EOM:general}
\Eeq
where $\tilde Q_{I \mu}$ represents the energy-momentum transfer 
term for the component $I$, $\tilde u^\mu$ is the 4-velocity of the
whole matter system, and $\tilde Q_I:=-\tilde u^\mu\tilde Q_{I\mu}$.
Because of the conservation of the total energy-momentum,
$\tilde Q_{I \mu}$ satisfies
\Beq
\sum_I\tilde Q_{I\mu}=0.
\label{EMTransferRate:constraint}\Eeq

For the scalar perturbation, the energy-momentum tensor and the
energy-momentum transfer vector of each individual component are expressed
as
\Beqr
&& \tilde T^{~0}_{I 0}=
  -(\rho_I+\delta\rho_I Y),\\
&& \tilde T^{~0}_{I j} = 
  a(\rho_I+p_I)(v_I-B)Y_j,\\
&& \tilde T^{~j}_{I k} = 
 (p_I\delta^j_k + \delta p_I\delta^j_k + \Pi_I Y^j_k),\\
&& \tilde Q_{I 0} =
 - [Q_I +(Q_I A + \delta Q_I)Y],\\
&& \tilde Q_{I j} =
 a [Q_I (v-B) + F_{c I}] Y_j,
\Eeqr
where $\rho_I$, $p_I$ and $Q_I$ are the background values of the
energy density, the pressure and the energy transfer of the component
$I$, respectively.  In terms of these quantities the background part
of \Eq{EOM:general} is written as
\Beq
 \dot \rho_I=-3 h_I(1-q_I)H, 
\Eeq
where 
\Beq
 h_I := \rho_I+p_I, \quad 
q_I := Q_I / (3 H h_I).
\Eeq

As in \S2, we construct from the density perturbation $\delta\rho_I$,
the velocity perturbation $v_I$, the isotropic stress perturbation
$\delta p_I$ and the energy transfer perturbation $\delta Q_I$ the
following four gauge-invariant combinations:
\Beqr
&& \rho_I \Delta_{g I}:= \delta\rho_I + 3(\rho_I+p_I)\R,\\
&& V_I:=v_I-{a\over k}\dot H_T,\\
&& p_I \Gamma_I:=\delta p_I - c_I^2\delta \rho_I,\\
&& Q_I E_{g I}:=\delta Q_I - {\dot Q_I \over H}\R
\Eeqr
where $c_I^2=\dot p_I/\dot \rho_I$.  The anisotropic stress
perturbation $\Pi_I$ and the momentum transfer perturbation $F_{cI}$
of the component $I$ are gauge-invariant by themselves.

>From the relation $\tilde T_{\mu \nu}= \sum_I \tilde T_{I \mu \nu}$
the variables for the whole system are expressed in terms of those for
each component as
\Beqr
&& \rho = \sum_I \rho_I,\quad
p = \sum_I p_I,\quad
h = \sum_I h_I,\\
&& \rho \Delta_g = \sum_I \rho_I \Delta_{g I},\quad
h V = \sum_I h_I V_I,\quad
p \Gamma = \sum_I p_I \Gamma_I + p \Gamma_{rel},\quad
 \Pi = \sum_I \Pi_I,
\Eeqr
where $p \Gamma_{rel} = \sum_I (c^2_I-c^2_s) \rho_I \Delta_{g I}$.
Further \Eq{EMTransferRate:constraint} gives the constraints
\Beq
\sum_I Q_I = 0,\quad
\sum_I Q_I E_{g I}=0,\quad 
\sum_I F_{c I}=0.
\Eeq

In terms of these gauge-invariant variables the perturbed 
equations of motion \Eq{EOM:general} are written as
\Beqr
& \tilde \nabla_\nu \tilde T^\nu_{I 0}= \tilde Q_{I 0}:
& \Tdot{( \rho_I \Delta_{gI} )} + 3 H  \rho_I \Delta_{gI} 
 + {k^2 \over a^2 H} h_I (\Phi - Z_I) 
 + 3 H ( p_I \Gamma_I + c^2_I \rho_I \Delta_{gI}) 
\nonumber \\
&& \quad = Q_I {\cal A} + Q_I E_{gI},
\label{EOMTimeComponent}\\
& \nabla_\nu T^\nu_{I i}= Q_{I i}:
& k \left[ \Tdot{\left({h_I Z_I \over H}\right)}
 + 3 h_I Z_I + h_I {\cal A}  
 + p_I\Gamma_I + c^2_I \rho_I\Delta_{gI} 
 - {2 \over 3} \Pi_I \right] 
\nonumber \\
&& \quad = - a F_{cI} + k {Q_I \over H} Z,
\label{EOMSpaceComponent}
\Eeqr
where
\Beq
 Z_I:=\Phi-{aH\over k}V_I=\R - {aH\over k}(v_I-B).
\Eeq
In deriving these equations from the equations of motion
\Eq{EOM:general}, we have not multiplied any power of $k$. Hence
putting $k=0$ in these equations yields the equations for the exactly
homogeneous perturbations.

Now let us examine the relation of the $k\tend0$ limit of these
equations and the corresponding equations for the exactly homogeneous
perturbations with $k=0$. First note that $\tilde p_I$ and $\tilde
Q_I$ depend on the metric and the matter variables through their
scalar combinations.  In particular the 4-velocities appear in the
form $\tilde u_I^\mu\tilde u_{J\mu}$ and/or the scalar combinations of
$\tilde u_I^\mu$ and its covariant derivatives. In the linear
perturbation the terms proportional to $v_I$ produced from them all
come with positive powers of $k$. This implies that $\Gamma_I$ and
$E_{gI}$ are written as linear combinations of $\A$, $k^2\Phi$,
$\Delta_{gI}$, $k^2Z$ and $kV_{IJ}=k(v_I-v_J)$ with coefficients which
do not contain negative powers of $k$. Hence the $k\tend0$ limit of
\Eq{EOMTimeComponent} and the corresponding $k=0$ equation coincide
with each other under the condition $k^2\Phi\tend0$.

On the other hand, \Eq{EOMSpaceComponent}, which vanishes identically
for $k=0$ due to the vector origin of $F_{cI}$, gives a non-trivial
equation in the $k\tend0$ limit.
If we require the condition 
\Beq
 \Pi_I\tend 0,
\Eeq
it is written as
\Beq
 \Tdot{\left({h_I Z_I \over H}\right)}
 + 3 h_I Z_I + h_I {\cal A}  
 + p_I \Gamma_I + c^2_I\rho_I\Delta_{gI} 
 = {Q_I \over H} Z + \lim_{k\tend0}F_{c I}/k.
\Eeq
Since $Z$ is related to $\A$ by \Eq{AdditionalConstraint}, this
equation can be regarded as one determining the $k\tend0$ limit of
$Z_I$, or $\lim_{k\tend0}(v_I-B)/k$, which has no relation to the
quantities describing homogeneous perturbations, from a solution for
$\A$ and $\Delta_{gI}$ to the perturbation equations with $k=0$(recall
the argument on $\Gamma_I$). This is in accordance with the general
argument in \S3.

In general this gives a set of coupled equations for $Z_I$. However,
if the condition 
\Beq
F_{cI}/k \tend 0
\label{AsymptCond:F}\Eeq
is satisfied, it can be explicitly integrated to give
\Beq
 h_I Z_I = {H \over a^3} \left[
 C_I - \int_{t_0} dt a^3 
( h_I {\cal A} + p_I  \Gamma_I + c^2_I\rho_I \Delta_{gI}
 - 3 h_I q_I Z ) \right],
\label{integralwrtZI}
\Eeq
where $C_I$ is an integration constant and $t_0$ is an initial time.
In particular, for the universal adiabatic modes given in \S4, this
gives 
\Beq
 h_I Z_I = {H \over a^3}\left[C_I 
+ \lambda \left\{{a^3 \over H}h_I-\left({a^3 \over H} h_I\right)_0
          \right\}\right],
\Eeq
where the subscript $0$ denotes the value at $t=t_0$. Here note that
\Eq{AsymptCond:F} is a rather strong condition in realistic situations
where $F_{cI}$ is usually proportional to the relative velocities
$V_{IJ}$.

As discussed in \S3, in order for this $k\tend0$ limit solution to
correspond to a $k=0$ solution, the consistency conditions (\ref
{AsymptCond2}) and (\ref {AsymptCond3}) should be satisfied. Under the
requirement (\ref{AsymptCond:F}), these conditions are simply
reduced to the asymptotic condition
\Beq
 k C_I(k) \tend 0.
\Eeq
Further, $Z_I$ is also restricted by the $k\tend0$ limit equation
(\ref{AdditionalConstraint}). Since $Z$ is expressed as
\Beqr
 h Z &=& \sum_I h_I Z_I
\nonumber \\
&=& {H \over a^3}\left[
 \sum_I C_I + \int_{t_0} dt a^3
                 \left(
 - h {\cal A} - p \Gamma - c^2_s \rho \Delta_g
                 \right)                 
                \right]
\nonumber \\
&=& {H \over a^3}\left[ \sum_I C_I 
- {2 \over 3} \left\{ {a^3 \over H} \rho {\cal A}
     - \left({a^3 \over H} \rho {\cal A}\right)_0
              \right\}\right],
\Eeqr
from \Eq{HamiltonianConstraint} and \Eq{PertEnergyEq1},
this condition is expressed as
\Beq
 \sum_I C_I 
 + {2 \over 3}\left({a^3 \over H}\rho{\cal A}\right)_0=0.
\label{AdditionalConstraint:PF}\Eeq
This is just the equation (\ref{AdditionalConstraint}) at the initial
time $t=t_0$.

For a $N$-component perfect fluid system, the perturbation of the
exactly homogeneous solutions generates $N$-independent solutions for
$\Delta_{gI}$ and $\A$, which satisfy the $k\tend0$ limit of
\Eq{EOMTimeComponent} and the Einstein equations. Each of these
solutions in turn determines $Z_I$ through \Eq{integralwrtZI} with
$N-1$ independent integration constants satisfying the condition
(\ref{AdditionalConstraint:PF}).  By this procedure we obtain $2N-1$
independent solutions to the whole perturbation equations in the
$k\tend0$ limit. Thus, by adding the universal adiabatic decaying mode
corresponding to the trivial homogeneous perturbation, we can obtain
all the $2N$ independent $k\tend0$-limit solutions for this system
from the exactly homogeneous solutions.

\section{Summary and Discussions}

In this paper we have investigated the relation between the $k\tend0$
limit of solutions to the perturbation equations with $k\not=0$ and
the exactly homogeneous perturbations ($k=0$) for quite general
multi-component systems on a spatially flat Robertson-Walker universe. 
The main result is summarized as follows.

First for the case in which the fundamental variables describing
cosmic matter contain a dynamical vector field, for any solution to
the perturbation equations with $k=0$ there exists a solution to the
perturbation equations with $k\not=0$ which coincides with the former
in the $k\tend0$ limit. On the other hand if cosmic matter consists
only of scalar fields, such a correspondence holds if and only if the
exact homogeneous perturbation satisfies an additional constraint
corresponding to the momentum constraint which is expressed by a
linear equation on the initial condition for the homogeneous
perturbation. This implies that for the $N$-component scalar field
system one can obtain the $k\tend0$ limit of $2N-1$ independent
solutions to the perturbations equations if one knows all the
solutions to the homogeneous background equations.

We have also shown that from trivial exactly homogeneous perturbations
one can always construct two adiabatic solutions to the perturbation
equations in the $k\tend0$ limit which can be explicitly expressed in
terms of integrals of background quantities, irrespective of the
content and the interactions of cosmic matter. One of them
corresponding to the vanishing homogeneous perturbation represents a
decaying mode for which the Bardeen parameter vanishes. The other
corresponds to a simple constant scaling of the scale factor and gives
a growing mode whose Bardeen parameter is a non-vanishing
constant. This constant scaling is not a gauge transformation, but an
extra symmetry of the background equations which holds only for
spatially homogeneous Robertson-Walker universes. Thus this extra
symmetry is the hidden reason why exact solutions in the $k\tend0$
limit have been so far found for various systems in spite of the
non-trivial structure of their perturbation equations.

For the multi-component scalar field system these universal adiabatic
modes and the solutions obtained from the non-trivial exact
homogeneous perturbations exhaust all the solutions to the
perturbation equations in the $k\tend0$ limit. Hence the evolution of
perturbations on superhorizon scales can be determined only from the
knowledge on the homogeneous solutions to the background equations in
the lowest-order approximation.  However, if one wants to know the
higher-order correction in $k$, one must solve the perturbation
equations $k\not=0$ iteratively \cite{Kodama.H&Hamazaki1996}. In this
procedure one needs the $k\tend0$ limit of all the solutions to the
perturbation equations for the gauge-invariant variable describing the
perturbation of the scalar fields. As we have shown, this information
can be also obtained from the exactly homogeneous solutions to the
perturbation equations.  Thus the behavior of superhorizon
perturbations of this system can be determined with any accuracy from
the knowledge on the homogeneous solutions.

This argument can be directly extended to the scalar field system 
coupled with radiation, for which all the $k\tend0$ limit of the
solutions to the perturbation equations are directly determined from
the homogeneous background solutions. This correspondence may provide 
a very powerful method to analyze the evolution of perturbations 
during reheating in the inflationary models because the structure of
the equations for the homogeneous scalar fields decaying to radiation 
is much simpler than that for the perturbation equations with energy
transfer terms.

\section*{Acknowledgments}

The authors would like to thank K. Nakao, Y. Nambu, M. Sasaki,
T. Tanaka, A. Taruya and K. Tomita for valuable discussions and
comments, without which this paper would not have emerged.
T. H. would like to thank Prof. H. Sato for continuous encouragements.
H. K. is supported by the Grant-In-Aid of for Scientific Research (C)
of the Ministry of Education, Science, Sports and Culture in
Japan(05640340).

%\bibliographystyle{/home/usr2/kodama/tex/inputs/jpap}
%\bibliographystyle{jpap}

%\bibliography{k0limit}

\end{document}